\documentclass[12pt]{article}   
 \usepackage{psfig,times,fancyheadings}
        
\begin{document}
\thispagestyle{empty} 


 \lhead[\fancyplain{}{\sl }]{\fancyplain{}{\sl }}
 \rhead[\fancyplain{}{\sl }]{\fancyplain{}{\sl }}


\newcommand{\nc}{\newcommand}

\nc{\qI}[1]{\section{ {#1} }}
\nc{\qA}[1]{\subsection{ {#1} }}
\nc{\qun}[1]{\subsubsection{ {#1} }}
\nc{\qa}[1]{\paragraph{ {#1} }}

\nc{\qfoot}[1]{\footnote{ {#1} }}
\def\qL{\hfill \break}
\def\qpar{\vskip 2mm plus 0.2mm minus 0.2mm}

\def\qparr{ \vskip 1.0mm plus 0.2mm minus 0.2mm \hangindent=10mm
\hangafter=1}

\def\qdec#1{\par {\leftskip=2cm {#1} \par}}

\def\qdpt{\partial_t}
\def\qdpx{\partial_x}
\def\qddpt{\partial^{2}_{t^2}}
\def\qddpx{\partial^{2}_{x^2}}
\def\qn#1{\eqno \hbox{(#1)}}
\def\qds{\displaystyle}
\def\qal{\sqrt{1+\alpha ^2}}
\def\qw{\widetilde}


\null
\vskip 3cm

\centerline{\bf \LARGE Speculative trading:}
\vskip 0.5cm
 \centerline{\bf \LARGE the price multiplier effect}

\vskip 1cm
\centerline{\bf B.M. Roehner $ ^* $ }
\centerline{\bf L.P.T.H.E. \quad University Paris 7 }

\vskip 2cm

{\bf Abstract}\  During a speculative episode the price of an item jumps
from an initial level $ p_1 $ to a peak level $ p_2 $ before more or less
returning to level $ p_1 $. 
The ratio $ p_2/p_1 $ is referred to as the amplitude $ A $ of the peak. 
This paper shows that for a given market the peak amplitude is a linear function
of the logarithm of the price  at the beginning of the speculative
episode; with $ p_1 $ expressed in 1999 euros the relationship takes the form: 
 $ A=a\ln p_1 +b $; the values of the parameter $ a $ turn out to 
be relatively
independent of the market considered: $ a \simeq 0.5 $, the values of the
parameter $ b $ are more market-dependent, but are stable in the course of time
for a given market.
This relationship
suggests that the higher the stakes the more ``bullish'' the market becomes.
Possible mechanisms of this ``risk affinity'' effect are discussed. 

\vskip 1cm
  \centerline{\bf  27 September  1999 }

\vskip 1cm
{\bf PACS.}\ 64.60 Equilibrium properties near critical points - 
 87.23Ge Dynamics of social systems

\vskip 2cm
* Postal address: LPTHE, University Paris 7, 2 place Jussieu, 75005 Paris, France
\qL
\phantom{* }E-mail: ROEHNER@LPTHE.JUSSIEU.FR
\qL
\phantom{* }FAX: 33 1 44 27 79 90

\vfill \eject

\qI{Introduction}

Are the mechanisms of speculative trading basically the same in their 
different manifestations, that is to say whether one considers stocks,
property values, diamonds, futures contracts for commodities,  postage 
stamps, etc? Econophysicists are probably inclined to answer affirmatively and
to posit that all speculative markets are alike. D. Stauffer recently told
me that he had not read in the econophysical literature any statement to 
the contrary. Yet, economists generally hold the opposite view; as a matter
of fact most of them would even found that question somewhat weird. 
In any case
little statistical evidence has so far been produced in support of one 
or the other claim. One earlier attempt in that direction was a paper
that pointed out that the shape of the price peaks belong to the same
class of functions [11]. In the present paper
I analyze a specific feature of speculative trading which will be
referred to as the price multiplier effect and I show that it can be
observed in various speculative markets for which adequate data exist,
particularly in real estate, diamond and stamp markets. \qL
Before coming to this let us briefly discuss the following question: 
Why, instead 
of concentrating on just one market, is it important to consider different
speculative markets? After all understanding the stock market already poses
a formidable challenge and it could seem a good strategy not to disperse 
efforts. The point is that once we know that speculative attitudes are
basically the same in any market the model we will build for the stock market
will not be the same as if we had limited our knowledge to 
that market alone; common factors will be emphasized while idiosyncrasies 
will be discarded. 
As a matter of illustration remember that historically it made a big
difference to realize that the fall of an apple, the ``fall'' of the moon and
the rising/falling of the tides are different facets of the same phenomenon. 
Similarly if one can prove that the mechanisms of speculation are 
basically the same everywhere this has far reaching consequences. One of them
is the idea
that the phenomenon of speculative trading is a sociological rather than an
purely economic phenomenon. 

\qI{The price multiplier effect}
\qA{Position of the problem}
The crucial question regarding speculation can be formulated as follows:
Suppose you have bought diamonds worth one million euros two years ago, assume
that in the meanwhile their price has risen by 360\%, but that in recent
weeks there was a sudden 25\% fall. Will you sell, or keep your holding until
the price goes up again or even buy more  diamonds? In the second and last 
cases speculation will be fueled and will continue at least for some time, with
the result that prices will rise to even higher levels. 
On
the contrary if you sell, and if other traders react similarly, the bubble
is likely to burst. \qL 
The relevance of the previous question is further emphasized by the following
observations (i) The above question is not a mere ``Gedanken experiment'', it
is based on the speculative episode that seized the diamond market in the
late 1970s; naturally, it 
can be rephrased in similar terms
for any other speculative item. (ii) Economists would claim that the above
question can be answered by relying on the standard principle of expected
revenue maximization. While it can be argued that future revenue of a stock 
is to some extent determined by the growth perspectives of the company, this 
is much more  questionable for diamonds. As a matter of fact it is precisely
to stress the fragility of the fundamentalists' argument that we picked up that
example. (iii) Numerous computer simulations of the stock market
have been proposed either by econophysicists or by economists; in that respect
one should mention the following works:[2,3,5,6,7,8,10,13]. Often
these simulations are based on astute mechanisms such as minority games or  
Ising type interactions. Yet, one would be on much firmer ground if these
simulations could be built on realistic attitudes at the micro-economic level. 
Models in statistical physics are successful to the extent that their 
assumptions about atomic or molecular interaction are no too unrealistic. \qL
To the above question the present paper provides the following answers (i)
The behavior of a speculator $ (A) $ who speculates on items worth 10,000 euros 
is not qualitatively different from that of a speculator $ (B) $ who speculates
(on the same market) on items worth 100 euros. This is reflected in the 
fact that the curves in Fig.1 have the same shape: same timing, same relaxation
time; only the amplitudes are different. (ii) Speculator $ A $ is significantly
more ``bullish'' than speculator $ B $. More specifically the amplitude
of the bubble (defined as the ratio between peak and initial price) is more or
less proportional to the initial price. \qL
At first sight the first point could seem fairly evident. It would 
be indeed for items whose prices differ only slightly, but for items whose
price differ by a factor of 100 it is no longer obvious that the attitude of the 
speculators should be the same. Speculator $ A $ is probably a professional 
while speculator $ B $ is likely to be a mere amateur. The second point means
that speculation will be stronger for 2-carat diamonds than for 0.5-carat
diamonds, for 5-room flats than for one-room apartments. How much stronger
will it be? This question is answered in the next paragraph. 

\qA{Statistical evidence}
\qun{Qualitative evidence}
Fig.1a,b,c,d presents the multiplier effect for four different speculative
bubbles. Each graph refers to items whose prices are markedly different, the
upper curve corresponding to the most costly item. The first figure concerns
the price of one-carat diamonds; as one knows the price of a diamond of
given size depends upon its color and number of flaws. Colorless diamonds are
the most costly, they are said to be of class D; classes ranging from F to 
Z correspond to diamonds of decreasing quality. The upper curve 
in Fig.1a corresponds to class D, while the lower is for class G. In 
normal conditions, that is to say before and after the bubble, there is a
ratio two between the prices of D and G diamonds. For these diamonds the 
peak amplitudes  (i.e. peak price / initial price) are 6.1 and 5.0 
respectively. \qL
Fig.1b describes a speculative bubble for property values in Paris. What is
shown is the price per square-meter of apartments in the 7th (one of the 
most expensive) and 19th (one of the less expensive) districts respectively. 
Again the amplitude of the peak is larger for the most expensive item: 2.83
against 1.95.  \qL
Fig.1c,d concern postage stamps. For our purpose postage stamps are of 
particular interest because their value range form a fraction of euro to several
thousand euros. During World War II there was a strong speculative bubble in 
France. Again we verify that the amplitude of the peak is larger for the item 
having the highest price; the figures are 5.6 and 1.9 respectively. 
Fig.1d refers to a speculative bubble for British stamps; again the most 
costly have the largest peak amplitude: 4.9 for the 5,000 franc stamp against
2.2 for the 275 franc stamp.  In this case we included an item for which 
there seems to be no speculative bubble at all. As a matter of fact in 
stamp catalogues one reads that speculation only concerned stamps with
a high face value. One may wonder why. In our explanatory framework the 
matter becomes simple: in fact speculation also affected the stamps 
with a low face value but these stamps being much cheaper their
price peak was small to the point of being 
at the same level as the average price. In the next
paragraph we will see that this explanation is not only qualitatively
satisfactory but although quantitatively correct. 

\qun{Relationship between price and peak amplitude}
From the above examples it is clear that peak amplitude increases 
with (initial) price. However to get a more precise idea of that
relationship one needs a statistical analysis of a larger sample of
cases. The corresponding data are summarized in Appendix A. 
With $ A $ denoting the peak amplitude and $ p_1 $ the initial price of the 
item  the relationship can be written in the form:
$$ A = a\ln p_1 + b $$

To give to the values of $ b $ an intrinsic meaning one has to make
the convention that $ p_1 $ must be expressed in a fixed currency.
We made the choice to use euros (of January 1999) as our currency scale. 
To take an example the francs of 1984 used for apartment prices in Paris
were transformed in euros through the following steps:

Euro = (1/6.56)\ F1999 ; F1999=(982/1408)\ F1984

The first equality is the official exchange rate between a
franc of January 1999 and an euro of January 1999 while the second is
based on a standard annual price index series. \qL

A linear least square fit gives the following estimates.
For diamonds, due to a lack of data, it was not possible to perform a
fit for a larger number of cases than in Fig.1a

\vskip 1cm

\centerline{\bf Table 1\quad Estimates for the parameters  $ a $ and 
 $ b $}

$$ \matrix{
\hbox{Item} \hfill & n & a & b & r \cr
  & \hbox{Number of}  &  &  & \hbox{Coefficient of} \cr 
  & \hbox{cases}  &  &      & \hbox{correlation} \cr 
  &  &  &        \cr 
\hbox{Appartments in Paris} \hfill & 20 & 0.60 \pm 0.50 & -1.7 \pm 0.1 & 0.48 \cr
  &  &  &        \cr 
\hbox{French stamps} \hfill & 6 & 0.47 \pm 0.38 & 2.2 \pm 0.7 & 0.77 \cr
  &  &  &        \cr 
\hbox{British stamps} \hfill & 13 & 0.39 \pm 0.17 & 2.3 \pm 0.6 & 0.80 \cr
} $$

\vskip 1cm

Two observations are in order (i) In each case the correlation is 
significantly larger than zero (ii) The estimates for $ a $ and $ b $ are
fairly close. \qL

\qI{Conclusion}
We have shown (at least  for those markets for which data were available)
that the relative amplitudes $ A $ of speculative price peaks are larger
for more costly items ; more specifically the relationship can be 
written in the form: $ A=0.5\ln p_1 + b $. Why is this so? \qL
Different mechanisms can be imagined. We will in this empirical paper refrain
from proposing a detailed model; this will be done in a
subsequent paper. Nevertheless it can be of interest to review some of the
ideas on which such a model could be based. For the sake of illustration
let us consider for instance the real estate market. We assume that there
are two types of operators: (i) Residents who buy and sell apartments for
personal usage; we  call them users (ii) Speculators and property developers
who make money by selling and buying property. Such a distinction is in essence
similar to the one made between  ``fundamentalists'' and  ``noise traders'' in
the paper by Lux et al. ([6]). Now, it is not unreasonable to assume that 
having limited means the users will be deterred by too high prices. On the
contrary for someone who buys  in order to sell 6 months later the price 
makes little difference; only profit matters.  
For instance if the price of one-room apartments
doubles users would still be able to afford them; on the contrary if the
price of 5-room apartments doubles they will become far too expensive. 
Consequently, one can expect that for costly goods the proportion of
speculators in the market will be larger. The dynamic pricing behavior
of speculators being more aggressive one should not be surprised that 
these markets show greater amplification factors. \qL
It is possible to check that scenario empirically at least to some extent. 
In a separate paper
[12] we tried to estimate the proportion of speculators in different
districts of Paris; for instance for the two districts considered in Fig.1b
namely the 7th and the 19th one gets 18\% and 11\% respectively; 
for all the districts the percentage of speculators varies from 10\% 
(2th district) to 36\% (15th district).
To get
these estimates we used the fact that 
on average, i.e. for 
all the 20 districts, the proportion of speculators
is about 20\% (La Vie Fran\c caise
18 April 1998). The regression between peak amplitudes and fraction
of speculators $ f_s $ reads: $ A=1.02\ f_s +2 $, the correlation being
equal to 0.28. \qL
This test is not completely satisfactory because both the peak amplitudes
and the proportions of speculators varied within too narrow limits: from
1.95 to 2.83 for the amplitudes and from 0.10 to 0.36 for the proportion
of speculators. It will be the purpose of a subsequent paper to 
perform similar tests in other markets. The main difficulty in that matter
is to find adequate statistics. \qL
As a last point one could wonder what (if any) are the implications of
the multiplier effect for stock markets. In this case one should of
course not reason in terms of share prices (these are of the order of
60 euros and are not very different from one stock to another) but in 
terms of share packages. For example, some small investors, called
``odd lotters'' in the jargon of finance, trade small portions of
stocks, typically under one hundred at a time. Around 1955 odd lotters
were known to be responsible for around 15\% of stock trading on the 
New York Stock Exchange; today they account for less than 1\% [15]. 
In order to test the price multiplier effect one would need more
detailed statistics about the size of transactions on given stocks. 
Once again finding adequate data turns out to be a major obstacle. 

\qpar

{\bf Acknowledgment} \ I am indebted to Mr. Nicolas Vuillet, diamond
trader at Paris, for very interesting and
stimulating discussions; they provided one of the starting points for the
writing of this paper. Furthermore I am grateful to the referee thanks to 
whom the paper has been markedly improved. 

\vfill \eject

\appendix

\qI{Appendix A: Statistical data}
In this appendix we give the detailed data for each of the cases 
considered above.
The peak amplitudes
are always computed from deflated prices. 

\qA{Real estate bubble in Paris}
Apartment prices in 1984 are expressed in thousand 
French francs per square meter. 

$$ \matrix{
\hbox{District number} \hfill & (6) & (16) & (7) & (8) & (5) & (15) & (4) &
 (14) & (1) & (17) \cr
\hbox{Price in 1984} \hfill & 11.6 & 11.1 & 10.3 & 10.1 & 9.71 & 9.60 & 9.45 &
 8.81 & 8.03 & 7.91 \cr
\hbox{Peak amplitude} \hfill & 2.47 & 2.51 & 2.83 & 2.78 & 2.38 & 2.02 & 2.43 &
 2.03 & 2.71 & 2.31 \cr
 &  &  &  &  &  &  &  &  &  &   \cr
\hbox{District number} \hfill & (12) & (2) & (13) & (3) & (19) & (11) & 
(9) & (20) & (18) & (10) \cr
\hbox{Price in 1984} \hfill & 7.85 & 7.37 & 7.19 & 6.99 & 6.49 & 6.45 & 
6.35 & 6.11 & 5.90 & 5.51   \cr
\hbox{Peak amplitude} \hfill & 1.96 & 2.19 & 2.18 & 2.63 & 1.95 & 2.16 & 2.36 & 
2.05 & 2.05 & 2.27 \cr
} $$

\qA{French stamps}
The stamp identification numbers refer to the C\' er\` es catalogues. The 1938
prices are expressed in current French francs.

$$ \matrix{
\hbox{Stamp number} \hfill & (2) & (31) & (30) & (32) & (11) & (16) \cr
\hbox{Price in 1938} \hfill & 5,000 & 400 & 350 & 250 & 60 & 9 \cr
\hbox{Peak amplitude} \hfill & 5.56 & 2.78 & 2.64 & 4.07 & 3.70 & 1.93 \cr
} $$

\qA{British stamps}
The stamp identification numbers refer to the Yvert 
and Tellier catalogues. The 1975
prices are expressed in current French francs.

$$ \matrix{
\hbox{Stamp number} \hfill & (90) & (46) & (89) & (156) & (105) & (183) &
 (155) \cr
\hbox{Price in 1975} \hfill & 13,500 & 8,000 & 6,000 & 2,250 & 1,500 & 1,400 &
 300 \cr
\hbox{Peak amplitude} \hfill & 4.87 & 2.84 & 5.47 & 4.33 & 5.05 & 2.74 &
 5.44 \cr
 &  &  &  &  &  &  &  \cr
\hbox{Stamp number} \hfill & (286) & (238) & (355) & (239) & (106) & (512) & 
 & \cr
\hbox{Price in 1975} \hfill & 275 & 90 & 3 & 1.75 & 1.50 & 1.35 & \cr
\hbox{Peak amplitude} \hfill & 2.19 & 2.85 & 0.36 & 0.76 & 0.86 & 1.31 & \cr 
} $$

\vfill \eject

\centerline{\bf \Large References}

\vskip 1cm

\qparr
(1) BERTERO (E.M.) 1987: Speculative bubbles and the market for precious 
metals. PhD thesis. London.

\qparr
(2) BOUCHAUD (J.-P.), POTTERS (M.) 1997: Th\'eorie des risques financiers.
Alea-Saclay, Eyrolles. Paris.

\qparr
(3) CALDARELLI (G.), MARSILI (M.), ZHANG (Y.-C.) 1997:  Europhysics Letters 40,479.

\qparr
(4) FAY (S.) 1982: The great silver bubble. Hodder and Stoughton. London. 

\qparr
(5) FEIGENBAUM (J.A.), FREUND (P.G.O.) 1996: International Journal of Modern Physics B 10,3737.

\qparr
(6) LUX (T.), MARCHESI (M.) 1999:  Nature 397, 498.

\qparr
(7) MANTEGNA (R.N.), STANLEY (H.E.) 1997: Physica A 239,255.

\qparr
(8) MANTEGNA (R.N.), STANLEY (H.E.) 1999: Scaling approach in finance. Cambridge
University Press. Cambridge (in press). 

\qparr
(9) MASSACRIER (A.) 1978: Prix des timbres-postes fran\c cais classiques de
1904 \`a 1975. A. Maury. Paris. 

\qparr
(10) OLIVEIRA (S.M. de), OLIVEIRA (P.M.C. de), STAUFFER (D.) 1999: Evolution,
money, wars and computers. Teubner. Stuttgart. See especially chapter 4
about stock market models. 
 
\qparr
(11) ROEHNER (B.M.), SORNETTE (D.) 1998: The European Physical Journal B 4, 387.

\qparr
(12) ROEHNER (B.M.) 1999: Regional Science and Urban Economics 29,73.

\qparr
(13) SORNETTE (D.), JOHANSEN (A.) 1997: Physica A 245,411.

\qparr
(14) STAUFFER (D.), OLIVEIRA (P.M.C.), BERNARDES (A.T.) 1999: International
Journal for Theoretical and Applied Finance 2,83.

\qparr
(15) TVEDE (L.) 1990: The psychology of finance. Norwegian University Press. 
Oslo.

\vfill \eject

Figure captions
\qpar

{\bf Fig.1a \quad Speculative bubble for polished diamonds.} 
Solid line: one carat, D clarity; broken line: one carat, G clarity; the price
data are deflated prices on the Antwerp market. The two figures under the
title give the prices 
of both items at the beginning of the bubble. {\it Sources: 
The Economist, Special report No 1126.}

{\bf Fig.1b \quad Speculative bubble for property values (Paris).} 
Solid line: price of apartments 
in one of the most expensive districts;
broken line: price of apartments in one of the less expensive districts;
the price data are deflated prices
per square meter. The two figures under the title give the prices of 
both items at the beginning of the bubble. {\it Source: Chambre des Notaires.}

{\bf Fig.1c \quad Speculative bubble for French stamps.} 
Solid line: price of one of the most expensive stamps; broken line: price 
of one of the cheapest stamps. The price data are deflated prices. The two
figures under the title give the prices of both stamps at the beginning of the bubble. 
{\it Source: Massacrier (1978).}

{\bf Fig.1d \quad Speculative bubble for British stamps.} 
Solid line: price of one of the most expensive stamps; broken line: price 
of a less expensive stamp; dotted line: price of one of the cheapest stamps; the 
price data are deflated prices. The three figures under the title give the
prices of the three stamps at the beginning of the bubble. {\it Source: Catalogue
Yvert and Tellier.}

\end{document}